\documentclass[traditabstract,a4,letter]{aa}  

\usepackage{epsfig}
\usepackage{graphicx}
\usepackage{epstopdf}
\usepackage{color}  
\usepackage{array}   
\usepackage{units}  
\usepackage[breaklinks,colorlinks, citecolor=blue]{hyperref}
\usepackage{natbib}  
\usepackage{multirow}   
\usepackage{amsmath,amssymb}  
\usepackage[english]{babel}
\usepackage[latin1]{inputenc}
\usepackage[T1]{fontenc}
\usepackage{longtable,lscape}
\usepackage{footnote}

\usepackage{txfonts}
\usepackage[toc,page]{appendix} 
\def\be{\begin{equation}}
\def\ee{\end{equation}}
\def\ba#1\ea{\begin{align}#1\end{align}}

\begin{document}

%%%%%%%%%%%%%

%%%%%%%%%%%%%%%
\title{Evolution of the mass-richness relation\\ for the redMaPPer catalog}
\author{G.Hurier\inst{1}}

\institute{
1 Centro de Estudios de F\'isica del Cosmos de Arag\'on (CEFCA),Plaza de San Juan, 1, planta 2, E-44001, Teruel, Spain
\\
\email{hurier.guillaume@gmail.com} 
}

\abstract{The accurate determination of the galaxy cluster mass-observable relations is one of the major challenge of modern astrophysics and cosmology. 
We present a new statistical methodology to constrain the evolution of the mass-observable relations.
Instead of measuring individual mass of galaxy clusters, we only consider large scale homogeneity of the Universe. 
In this case, we expect the present galaxy cluster mass function to be the same everywhere in the Universe.
Using relative abundance matching, we contraint the relation between the richness, $\lambda(z)$, and the expected present mass, $M(t_0)$, of galaxy clusters.
We apply this approach to the redMaPPer galaxy cluster catalogue in 10 redshift bins from $z=0.1$ to $0.6$.\\
We found that the $\lambda(z)$-$M(t_0)$ relation is not evolving from $z=0.1$ to $0.4$, whereas it starts to significantly evolve at higher redshift.
This results implies that the redMaPPer richness appears to be a better proxy for the expected present-day galaxy cluster mass than for the mass at the observational redshift.\\
Assuming cosmology and galaxy cluster mass accretion history, it is possible to convert $M(t_0)$ to the mass at the galaxy cluster redshift $M(t_z)$.
We found a significant evolution of the $\lambda(z)$-$M(t_z)$ over all the covered redshift range. Consequently, we provide a new redshift-dependent richness-mass relation for the redMaPPer galaxy cluster catalogue.
This results demonstrates the efficiency of this new methodology to probe the evolution of scaling relations compared to individual galaxy cluster mass estimation.}


\authorrunning{G.Hurier}
\titlerunning{A redshift dependent richness-mass relation for the redMaPPer catalog}

\maketitle
  
\section{Introduction}

Galaxy clusters are the largest gravitationally bound structures in the Universe.
Their observation is allowed by several probes: over-density of galaxies \citep{wen12,roz14}, weak lensing produced on background galaxies \citep{hey12,erb13}, X-ray emission produced by the hot gas of electrons within galaxy clusters through Bremsstrahlung radiation \citep{boh01}, and the thermal Sunyaev-Zel'dovich (tSZ) effect on the cosmic microwave background (CMB) produced by the same population of electrons \citep{sun72}.\\

Galaxy clusters are now widely used as a cosmological probe, using: galaxy cluster number count \citep{has13,rei13,planckszc,dah16}, angular power spectra in auto and cross-correlation \citep{planckszs,hur15a,hur15b,hur17c}, or higher order statistic such as the bispectrum \citep{hur17b}.\\

The main limitation in the exploitation of galaxy clusters as cosmological probe is the accurate determination of the mass-observable relations \citep{planckszc,hur15b,hur17b}.
The identification of galaxy clusters through over-densities of galaxies in photometric survey is the oldest detection method. However, the calibration of the relation between optical richness and the mass of galaxy clusters remain a complex problematic \citep{koe07,sar15,and16,far16,sim17,gea17,mur18}.\\
Several methods exist to determine galaxy cluster mass, including X-ray temperature \citep{ett02}, cluster velocity dispersion \citep{all11}, tSZ effect on the cosmic microwave background \citep{planckXI, planckXII}, and weak lensing \citep{sch05,mel15}.\\
All of these approach present advantages and drawbacks.
If the weak lensing allows to directly probe the total mass of galaxy clusters, its utilization for the mass-richness relation usually require large samples of galaxy clusters \citep[see e.g.,][]{gea17} or is limited to very massive galaxy clusters \citep[see a.g.,][]{and14}.\\
The calibration of the mass-richness relation is challenging, consequently the study of its evolution have been only marginally studied so far.
For very massive galaxy clusters, \citet{and14} didn't find any significant evolution of the mass-richness relation.\\ 

In this work, we use the redMaPPer \citep{roz14} catalogue to study the mass-richness relation evolution from z = 0.1 to 0.6. 
Instead of trying to infer the individual mass of each galaxy cluster, we present a new statistical method to produce a cosmology-dependent evaluation of the mass-observable relation evolution.\\
The paper is organized as follows: in Sect.~\ref{secmeth} we present the galaxy cluster mass function for the redMaPPer catalogue.
Then in Sect.~\ref{secres}, we present the statistical approach for the evaluation of the mass-richness relation evolution. Finally, in Sect.~\ref{secconcl}, we present the results for the redMaPPer galaxy cluster catalogue.

%made the make the hypothesis that galaxy clusters mass function should be the same at all position in the Universe. Assuming the cosmological parameters \citep{planckpar18} and the mass accretion history of galaxy clusters \citep{cor15}

\section{The redMaPPer galaxy cluster mass function}
\label{secmeth}

\begin{figure}[!h]
\begin{center}
\includegraphics[width=0.9\linewidth]{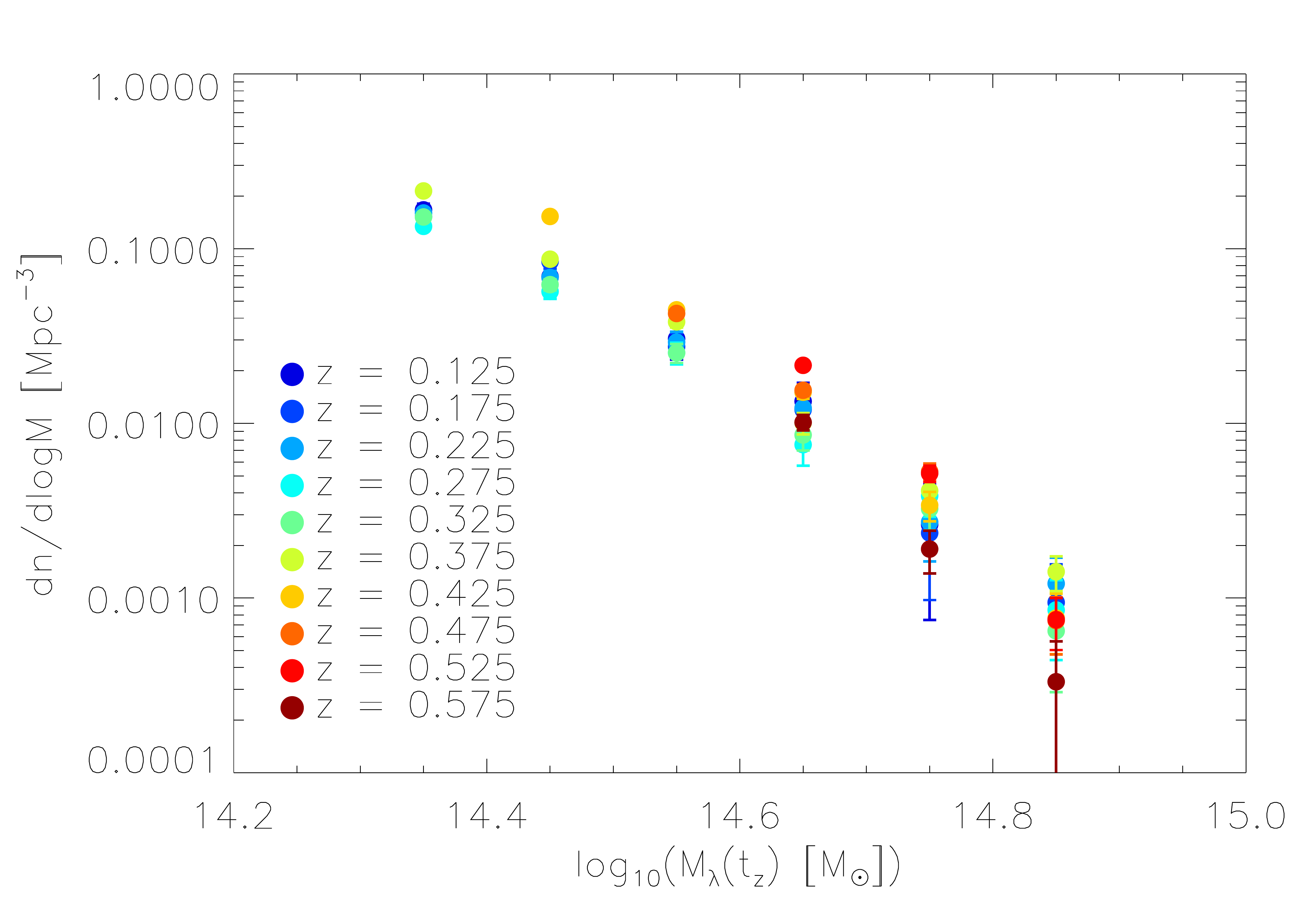}
\caption{Mass function of the redMaPPer galaxy clusters in 10 redshift bins from $z = 0.1$ to 0.6 as a function $M_\lambda(t_z)$.}
\label{dndmz}
\end{center}
\end{figure}

Under the assumption that the Universe is homogeneous, we expect the present day galaxy cluster mass function to be the same at all locations in the Universe.
We use the redMaPPer galaxy cluster catalogue \citep{roz14} composed of 26111 galaxy over-densities detected over $\sim 10500$ deg$^2$.
A key parameter of the redMaPPer catalogue is the richness, $\lambda$, which gives an estimate of the number of cluster member galaxies.
The number  of cluster members is expected to scale with the galaxy cluster total mass. Consequently, we expect to find a tight relation between $\lambda$ and the galaxy cluster mass.
This relation have been calibrated by previous studies. Using the mass-richness relation calibrated by \citet{gea17}, we convert redMaPPer richness, $\lambda$, into estimated mass, $M_\lambda(t_z)$.
Then, we constructed mass functions in 10 redshift bins from $z = 0.1$ to 0.6 as a function $M_\lambda(t_z)$, with mass bins of $\Delta {\rm log}_{10}M_{\lambda} = 0.1$. We estimated the uncertainties by assuming Poisson statistic.\\
Galaxy clusters are weighted by their corresponding purity and completeness \citep{roz14} accordingly to their richness and redshift. 
We compute the redshift dependent comoving volume assuming $Planck$-CMB best fitting cosmology \citep{planckpar18}.\\
In Table~\ref{tabcut} we present the minimum mass considered for the construction of the mass functions as a function of the redshift to avoid significant completeness issues at low-mass and high-redshift.

\begin{table}
\label{tabcut}
\caption{Redshift bin used for computing the galaxy cluster mass functions. The mass ${\rm log}_{10}M_{\lambda,{\rm min}}(z,t_z)$ gives the minimum mass considered for the mass function to avoid completeness issues, ${\rm log}_{10}M_{\lambda,{\rm min}}(z,t_0)$ provide the same information for the expected present-day galaxy cluster mass function assuming $Planck$-CMB cosmology and mass accretion history from \citep{cor15}.}
\center
\begin{tabular}{|c|cc|cc|}
\# & $z_{\rm min}$ & $z_{\rm max}$ & ${\rm log}_{10}M_{\lambda,{\rm min}}(t_z)$ & ${\rm log}_{10}M_{\lambda,{\rm min}}(t_0)$ \\
\hline
\phantom{0}1 & 0.10 & 0.15 & 14.3 & 14.4 \\
\phantom{0}2 & 0.15 & 0.20 & 14.3 & 14.4\\
\phantom{0}3 & 0.20 & 0.25 & 14.3 & 14.4\\
\phantom{0}4 & 0.25 & 0.30 & 14.3 & 14.4\\
\phantom{0}5 & 0.30 & 0.35 & 14.3 & 14.4\\
\phantom{0}6 & 0.35 & 0.40 & 14.3 & 14.5 \\
\phantom{0}7 & 0.40 & 0.45 & 14.4 & 14.6 \\
\phantom{0}8 & 0.45 & 0.50 & 14.5 & 14.7 \\
\phantom{0}9 & 0.50 & 0.55 & 14.6 & 14.8 \\
10 & 0.55 & 0.60 & 14.6 & 14.9\\
\end{tabular}
\end{table}

\noindent On Fig.~\ref{dndmz}, we present the 10 mass functions derived from $z = 0.1$ to 0.6. The mass functions are surprisingly nearly invariant with redshift, whereas in a standard cosmological scenario, we expect to see less massive galaxy clusters at high-redshift compared to low-redshift due to the growth of structures. In the present case, we even see that at $z \simeq 0.5$ galaxy clusters with ${\rm log}_{10}M_\lambda(t_z) = 14.7$ have a higher number density than clusters of the same mass at low redshift.\\
Using tSZ galaxy clusters \citep{PSZ2}, it has been recently shown that galaxy cluster number density is following the expected evolution with redshift \citep{hur19b}.\\
This result implies that the $\lambda(z)$-$M_{\lambda}(t_z)$ relation evolves significantly with $z$.\\
To illustrate this evolution, we compute the expected present-day mass of redMaPPer galaxy cluster assuming \citet{gea17} scaling relation, \citet{planckpar18} cosmology, and \citet{cor15} mass accretion history parametrization:
\begin{align}
\label{eqmah}
M(t_z) &= M(t_0) (1+z)^{\alpha f((t_0))}e^{-f(M(t_0))z}, \\
\alpha &= 1.686 \times (2/\pi)^{1/2} \frac{{\rm d}D}{{\rm d}z}|_{z=0} + 1, \\
f(M) &= 1/\sqrt{S(M/q) - S(M)},\\
q &= 4.137 \tilde{z}_f^{-0.9476},\\
\tilde{z}_f &= -0.0064 \times ({\rm log}_{10} M(t_0)^2) + 0.0237 \times ({\rm log}_{10} M(t_0)) \nonumber \\
& + 1.8837,
\end{align}
Where,
\begin{align}
S(M) =\frac{1}{2\pi^2} \int P(k) W^2(k,R) k^2 {\rm d}k,
\end{align}
with $P(k)$ the linear power spectrum of the matter distribution and $W(k,R)$ the Fourier transform of a top hat window function of radius $R$ corresponding to the mass $M$.\\

\noindent From this mass accretion history, we derive the expected present-day mass, $M_\lambda(t_0)$, for all redMaPPer galaxy clusters. We compute again the galaxy cluster mass function for the 10 redshift bins as a function of $M_\lambda(t_0)$.\\
\begin{figure}[!h]
\begin{center}
\includegraphics[width=0.9\linewidth]{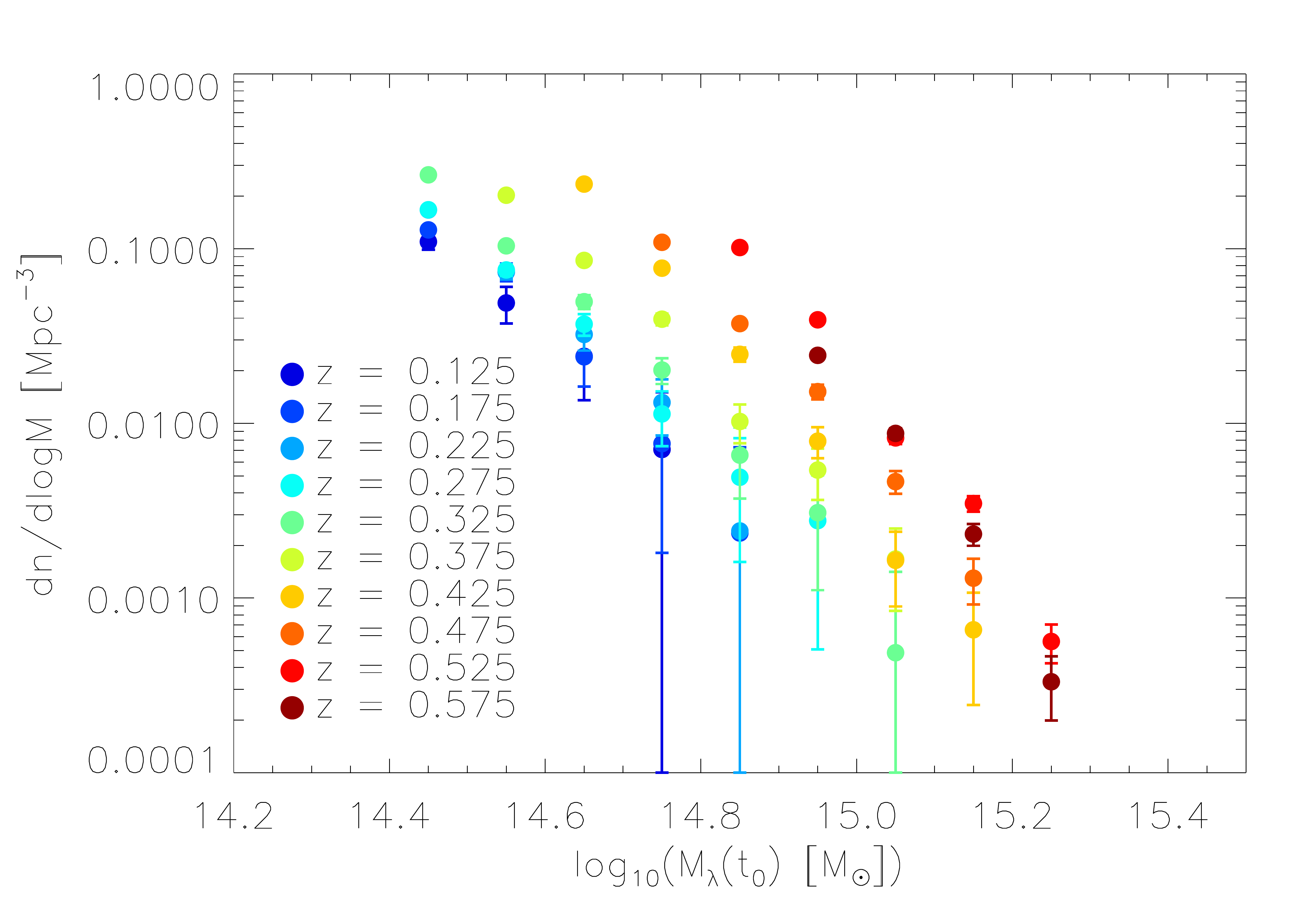}
\caption{Mass function of the redMaPPer galaxy clusters in 10 redshift bins from $z = 0.1$ to 0.6 as a function $M_\lambda(t_0)$.}
\label{dndm0}
\end{center}
\end{figure}
On Fig.~\ref{dndm0}, we present the redMaPPer galaxy cluster mass function for $M_\lambda(t_0)$. In an homogenous Universe, if the $\lambda$-$M$ relation was independent of the redhift, these mass-functions should be the same at all redshift.
However, we observe a significant evolution of the these mass functions, which implies that the $\lambda(z)$-$M_\lambda(t_z)$  scaling relation is strongly evolving with redshift and that this relation cannot be properly described with a redshift independent scaling relation.\\

\section{Relative abundance matching}
\label{secres}

Considering that the relation $\lambda$-$M_\lambda(t_z)$ from \cite{gea17} is already close to provide number density matching for the galaxy cluster mass functions, it implies that $\lambda(z)$ is in fact a good proxy for the expected $M_\lambda(t_0)$\footnote{The mass functions at various redshift are expected to match when expressed as a function of $M_\lambda(t_0)$.}.\\
In the following, we assume that the $\lambda(z)$-$M_\lambda(t_0)$ logarithmic scatter is invariant with redshift\footnote{In this case, the convolution of the mass function induced by the scatter is the same for all redshift bins.}.
We now aim at providing a measure of the $\lambda(z)$-$M_\lambda(t_0)$ relation evolution by performing relative abundance matching between the mass functions at various redshifts.\\
For each redshift bin we consider a scaling relation of the form,
\begin{align}
\label{scal}
{\rm log}_{10}M_\lambda(t_0) = \alpha(z) \, {\rm log}_{10}\left(\frac{\lambda(z)}{\lambda_0}\right) + {\rm log}_{10}M_0(z),
\end{align}
with $\lambda_0 = 40$, ${\rm log}_{10}M_0$ the pivot mass and $\alpha$ the slope of the mass-richness scaling relation.\\
We chose to express the richness-mass as a function of $M_\lambda(t_0)$ considering that this is the quantity we are matching when performing relative abundance matching.
It is worth noting that the relative abundance matching of the galaxy cluster mass function only allows to probe the scaling law evolution but does not allows an absolute calibration of this relation, unless we are assuming a theoretical mass function parametric form \citep[see e.g.,][]{tin08,wat12}.\\
Instead, we used the redshift bin $[0.40, 0.45[$ as the arbitrary reference mass function, and we adjust ${\rm log}_{10}M_0(z)$ and $\alpha (z)$ to obtain a match of the mass functions, ${{\rm d}n}/{{\rm d} M_\lambda}$, at all redshift.\\

\noindent Finally, we calibrate the $\lambda(z)$-$M_\lambda(t_0)$ relation by converting $M_\lambda(t_0)$ into $M_\lambda(t_z)$ assuming $Planck$-CMB cosmology, the mass accretion history from \citet{cor15}, and by imposing that the average $\lambda(z)$-$M_\lambda(t_z)$ scaling relation follows the result of \citet{gea17}.\\

\begin{table}
\label{tabpar}
\caption{Best fitting parameters for the $\lambda$-$M_\lambda(t_0)$ scaling relation as a function of galaxy cluster redshift. These results are cosmology dependent and assume \citet{planckpar18} best fitting cosmology, \citet{cor15} mass accretion history, and \citet{gea17} overall normalization.}
\center
\begin{tabular}{|c|cc|cccc|}
\# & $z_{\rm min}$ & $z_{\rm max}$ & $\alpha$ & $\Delta \alpha$ & ${\rm log}_{10}M_0$ & $\Delta {\rm log}_{10}M_0$ \\
\hline
\phantom{0}1 & 0.10 & 0.15 & 1.24 & 0.17 & 14.68 & 0.09 \\
\phantom{0}2 & 0.15 & 0.20 & 1.23 & 0.12 & 14.67 & 0.06 \\
\phantom{0}3 & 0.20 & 0.25 & 1.23 & 0.10 & 14.67 & 0.05 \\
\phantom{0}4 & 0.25 & 0.30 & 1.23 & 0.09 & 14.65 & 0.05 \\
\phantom{0}5 & 0.30 & 0.35 & 1.17 & 0.07 & 14.63 & 0.04 \\
\phantom{0}6 & 0.35 & 0.40 & 1.17 & 0.06 & 14.66 & 0.03 \\
\phantom{0}7 & 0.40 & 0.45 & 0.90 & 0.05 & 14.57 & 0.03 \\
\phantom{0}8 & 0.45 & 0.50 & 1.01 & 0.09 & 14.62 & 0.04 \\
\phantom{0}9 & 0.50 & 0.55 & 0.82 & 0.14 & 14.56 & 0.07 \\
10 & 0.55 & 0.60 & 0.80 & 0.19 & 14.50 & 0.08 \\
\end{tabular}
\end{table}

\section{Results}
\label{secconcl}

\noindent In Table~\ref{tabpar}, we list the values for the scaling relation parameters $\alpha(z)$ and ${\rm log}_{10}M_0(z)$.
We stress that this scaling laws yields for the mass at $z=0$\footnote{By performing abundance matching we are matching the number density of objects, a criteria which only apply if the masses are all expressed for the same moment of the Universe history.} and are therefore not providing directly the mass at the redshift of the clusters. Consequently, to derive the proper masses at the galaxy cluster redshift, $M_\lambda(t_0)$ has to be converted into $M_\lambda(t_z)$ using \citet{cor15} mass accretion history.\\

\begin{figure}[!h]
\begin{center}
\includegraphics[width=0.9\linewidth]{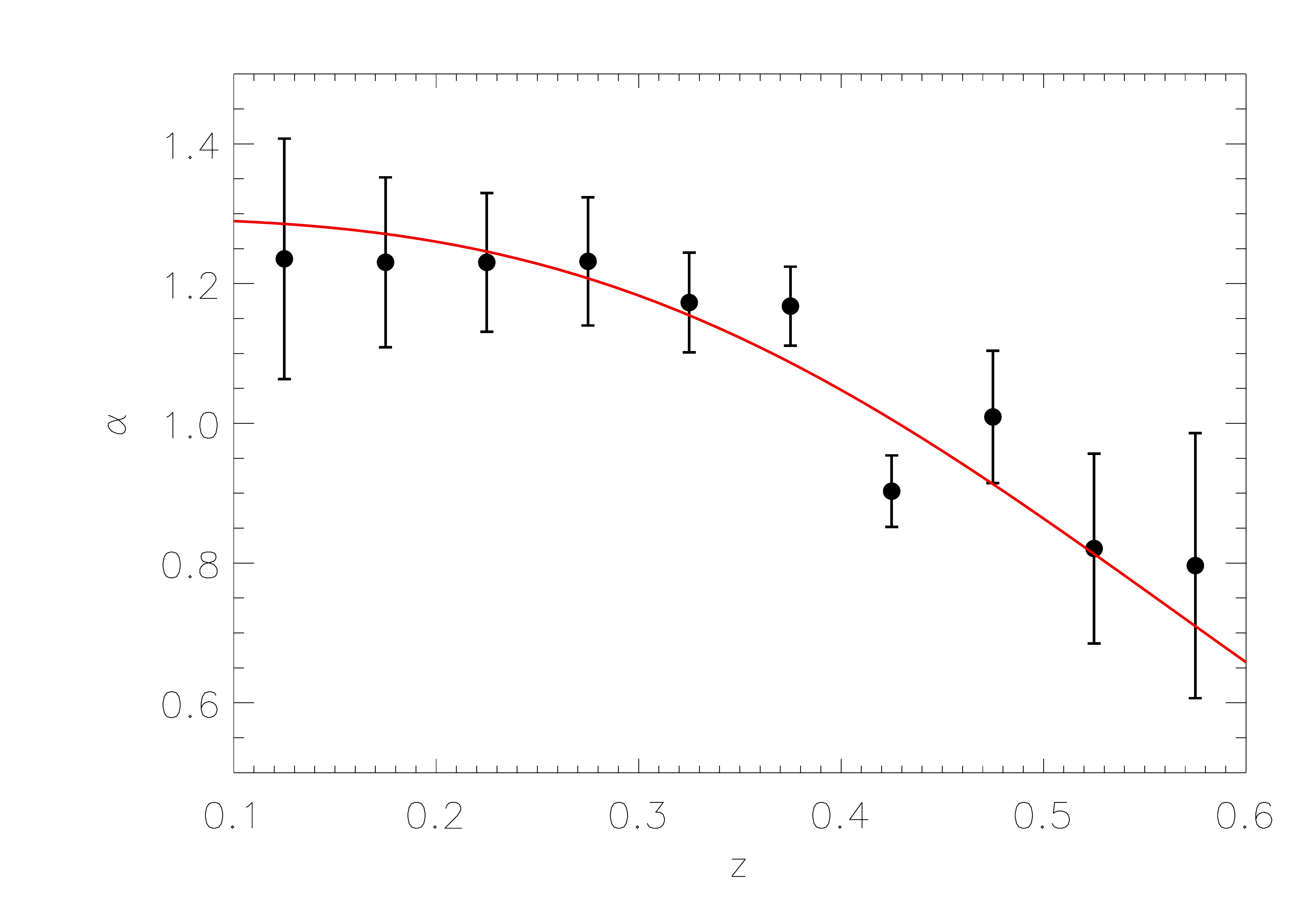}
\caption{Slope of the richness to present-day mass relation $\lambda(z)$-$M_\lambda(t_0)$ as a function of the galaxy cluster redshift. Black sample shows the values derived from relative abundance matching on the RedMaPPer galaxy cluster catalogue. The solid red curve shows the best adjustment of this evolution with a parametric function, see the text for more details.}
\label{figalpha}
\end{center}
\end{figure}

On Fig.~\ref{figalpha}, we show the best fitting values of $\alpha$ as a function of $z$ for the $\lambda(z)$-$M_\lambda(t_0)$ scaling relation (see Eq.~\ref{scal}). The stability of this relation from $z=0.1$ to 0.4 shows that redMaPPer richness $\lambda(z)$ is a better proxy of galaxy cluster expected present-day mass than of the galaxy cluster mass at their observational redshift, $M_\lambda(t_z)$. The rapid evolution of $\alpha$ at $z > 0.4$ is most likely induced by selection effect over the observed galaxies in SDSS survey.\\
We found that the redshift evolution of $\alpha$ is well fitted by the parametric function,
\begin{align}
\alpha = \frac{1.29}{(1+z^{3.04})^{3.52}},
\end{align}
which reproduce fairly the evolution of $\alpha$ for $z \in [0.1, 0.6[$. It is worth noting that $\alpha$ is always bigger than the slope of \citet{gea17} mass-richness relation. 
It is explained by the mass dependance of the mass accretion history that is not just affecting the scaling relation normalization but also the slope of the scaling relation.\\

\begin{figure}[!h]
\begin{center}
\includegraphics[width=0.9\linewidth]{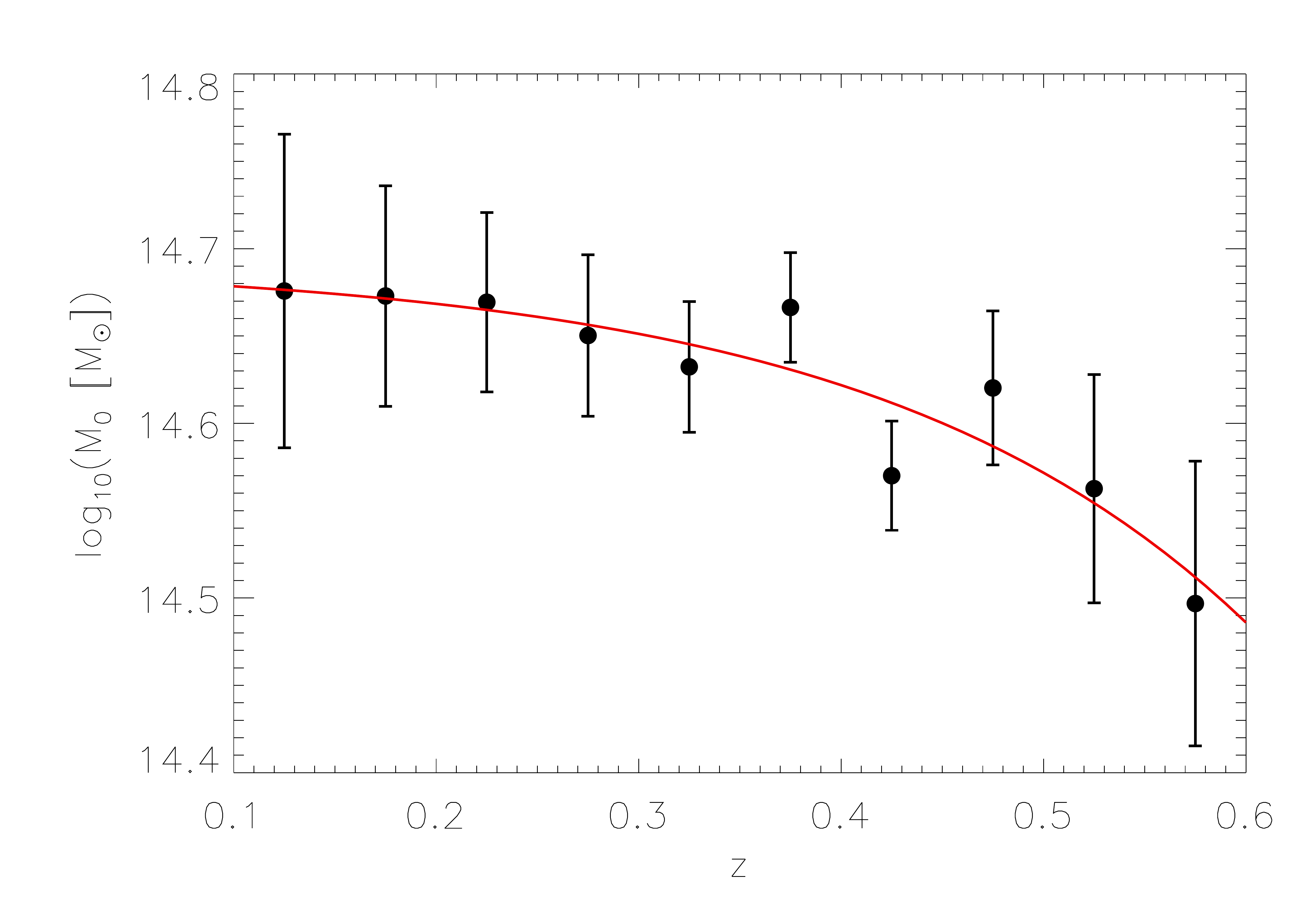}
\caption{Pivot mass of the richness to present-day mass relation $\lambda(z)$-$M_\lambda(t_0)$ as a function of the galaxy cluster redshift. Black sample shows the values derived from relative abundance matching on the RedMaPPer galaxy cluster catalogue. The solid red curve shows the best adjustment of this evolution with a parametric function, see the text for more details.}
\label{figlogm0}
\end{center}
\end{figure}

\noindent On Fig.~\ref{figlogm0}, we present the redshift evolution for the $\lambda(z)$-$M_\lambda(t_0)$ relation pivot mass, ${\rm log}_{10}M_0$.
We observe that the ${\rm log}_{10}M_0$ evolution is strongly correlated to the evolution of $\alpha$. This correlation is induced by the intrinsic correlation of the parametric form of the scaling relation\footnote{The amount of correlation depends on the choice of the pivot richness, $\lambda_0$.}.\\
More specifically, the evolution of the pivot mass as function of redshift shows that a non-evolving richness-mass relation will tend to over-estimate the mass of high-z galaxy clusters and under-estimate the mass of nearby objects, which leads to the mass functions presented on Fig.~\ref{dndm0}.\\
We found that the pivot mass evolution is well adjusted by the following parametric function
\begin{align}
{\rm log}_{10}M_0 = 14.69 \times \left[1-{\rm exp}(5.36\, z - 7.48)\right],
\end{align}
for $z \in [0.1,0.6[$.

\section{Conclusion}

We have presented a new cosmology-dependent statistical approach to evaluate the redshift evolution of mass-observable relations.\\
We applied this technic to the redMaPPer galaxy cluster catalogue to measure the evolution of the mass-richness relation.
Contrary to previous work \citep{and14}, we observe a clear evolution of the richness-mass relation. 
This results illustrates that redshift-independent mass-richness relations are not providing satisfying estimation of the galaxy cluster mass.
It also demonstrates the efficiency of relative abundance matching for the determination of mass-observable relation evolution when using large sample of galaxy clusters. \\

\noindent We found that the redMaPPer richness appears to be a better proxy for the expected present-day mass, $M_\lambda(t_0)$, than for the mass at the observation redshift, $M_\lambda(t_z)$.\\
Assuming \citet{planckpar18} cosmology, \citet{cor15} mass accretion history, and \citet{gea17} overall normalization we provide a new redshift evolving richness-mass relation for $M_\lambda(t_0)$.
Then, this relation can be converted into a $\lambda(z)$-$M_\lambda(t_z)$ relation using accretion history of galaxy clusters.  
We stress that the $\lambda(z)$-$M_\lambda(t_0)$ overall calibration is cosmology dependent as it assumes already a $M_\lambda(t_0)$ to $M_\lambda(t_z)$ conversion.
Whereas, the constraints on the evolution of the mass-richness relation only depend on cosmology through the considered redshift to comoving volume relation.\\

\section*{Acknowledgment}
\thanks{G.H. acknowledge support from Spanish Ministerio de Econom\'ia and Competitividad (MINECO) through grant number AYA2015-66211-C2-2. U.K. acknowledges support by the Israel Science Foundation (ISF grant No. 1769/15).}

\bibliographystyle{aa}
\bibliography{LM_evo}

\end{document}